\documentstyle[12pt,graphicx]{article}
\title{The hydraulic jump as a white hole}
\author{G.E. Volovik\\
Low Temperature Laboratory, 
Helsinki University of Technology\\
P.O.Box 2200, FIN-02015 HUT, Finland\\
and\\
L.D. Landau Institute for Theoretical Physics, 
 Moscow\\
}

\begin{document}
\maketitle
\abstract{ In the geometry of the circular hydraulic
jump, the velocity of the liquid in the interior region
exceeds the speed of  capillary-gravity waves  (ripplons), whose spectrum is 
`relativistic'  in the shallow water limit.
The velocity flow is radial and outward, and thus  the relativistic
ripplons cannot propagate into the interior region. In terms of the
effective 2+1 dimensional Painlev\'e-Gullstrand metric appropriate for
the propagating ripplons, the interior region  imitates the white hole.
The hydraulic jump represents the physical singularity at the white-hole
horizon. The instability of the vacuum in the ergoregion inside the
circular hydraulic jump and its observation in recent experiments on
superfluid $^4$He by   E. Rolley, C. Guthmann, M.S. Pettersen and C.
Chevallier \cite {Pettersen} are discussed.}
\vspace{5mm}

\section{Introduction}

Starting with the pioneering acoustic black hole  \cite{Unruh1981}, 
there appeared many suggestions to simulate the black and white holes in
various laboratory systems (see review paper  \cite{VisserReview} and
references therein). Here we discuss the most perspective analog, which
has been actually realized in recent experiments with superfluid $^4$He
\cite{Pettersen}: the circular hydraulic jump   in superfluid $^4$He
simulates the 2+1 dimensional white hole for the  surface waves
with ``relativistic''  spectrum in the shallow liquid.

 In Sec. \ref{SecEffectiveMetric} we discuss the effective space-time 
emerging for the surface waves -- ripplons -- in the shallow water limit.
In Sec. \ref{Instability} we introduce interaction of ripplons with the 
walls. The walls provide the absolute reference frame.  In the region
where the flow of the liquid with respect to this frame exceeds  Landau
critical velocity for ripplon radiation,  the surface of the liquid
becomes unstable. For the relativistic ripplons the boundary of this
region serves as analog of a black-hole or white-hole horizon. The
instability of the liquid towards generation of ripplons inside the
horizon 
 is the main mechanism of the decay of this 2+1 dimensional analog  of
the black or white hole. Similar instability of the vacuum inside the
astronomical black hole is possible.  In Sec. \ref{WH} we show that the
hydraulic jump is the realization of the white hole horizon for the 
relativistic ripplons  in normal  liquids,  In Sec. \ref{HJ} the
discussion is extended for the hydraulic jump in  superfluids in relation
to the recent experiment \cite {Pettersen}. Some open questions require
further investigations,

\section{Effective metric for ripplons}
\label{SecEffectiveMetric}

The general dispersion relation $\omega({\bf k})$ for ripplons --  the
waves on the surface of a  liquid -- is
\begin{equation}
M(k) (\omega- {\bf k}\cdot {\bf v})^2
=  \rho g+k^2\sigma ~.
\label{GeneralRipplonSpectrum}
\end{equation}
Here  $\sigma$ is the
surface tension; $\rho$ is mass density of the liquids;  $\rho g$
is the gravity force; and ${\bf v}$ is the velocity of the liquid along
the surface.  The quantity $M(k)$ is the
$k$-dependent  mass of the liquid which is forced into motion by the
oscillating surface:
\begin{equation}
M(k)={\rho \over k ~{\rm tanh}~kh}~, \label{Mass}
\end{equation}
where $h$ is the thicknesses of the layer of the liquid.

The spectrum (\ref{GeneralRipplonSpectrum}) becomes ``relativistic''  in
the shallow water limit $kh\ll 1$,
$k\ll k_0$:
\begin{equation}
(\omega- {\bf k}\cdot {\bf v})^2
=  c^2k^2 + c^2k^4\left(\frac{1}{k_0^2} -  \frac{1}{3} 
h^2\right)~~,~~c^2=gh~~,~~k_0^2=\rho g/\sigma ~.
\label{RelatSpectrum}
\end{equation}
If the $k^4$ corrections are ignored, the spectrum of ripplons in the
$k\rightarrow 0$ limit is described by the effective metric
\cite{SchutzholdUnruh} 
\begin{equation}
g^{\mu\nu}k_\mu k_\nu=0~~,~~k_\mu=(-\omega, k_x,k_y)~,
\label{EffectiveMetric}
\end{equation} 
with the following elements
\begin{equation}
g^{00}=-1~,~g^{0i}=-v^i~,~g^{ij}=c^2\delta^{ij}-v^iv^j~.
\label{EffectiveMetricContravariant}
\end{equation} 
The interval describing the effective 2+1 space-time in which  ripplons
propagate along geodesics  and the corresponding covariant components of
the effective metric are
\begin{equation}
ds^2=g_{\mu\nu}dx^\mu dx^\nu~~,~~g_{00}=-1
+\frac{v^2}{c^2}~,~g_{0i}=-\frac{v^i}{c^2}~,
~g_{ij}=\frac{1}{c^2}\delta_{ij}~.
\label{EffectiveMetricCovariant}
\end{equation}
As distinct from the original acoustic metric introduced by Unruh
\cite{Unruh1981}, here $c$ is the speed of gravity waves. It is
typically much smaller than the speed of sound, which allows us to avoid
different hydrodynamic instabilities inherent to the acoustic analogs of
the horizon.

The spectrum (\ref{GeneralRipplonSpectrum}) is valid for the prefect
fluid, where dissipation due to friction and viscosity is neglected, and
it must be modified when the dissipation is added.
For the ripplons propagating at the interface between two superfluids the
dissipation leads to a simple extra term on the right-hand side of
Eq.(\ref{GeneralRipplonSpectrum})
\cite{VolovikBook,Volovik}:
\begin{equation}
M(k) (\omega- {\bf k}\cdot {\bf v})^2
=  \rho g+k^2\sigma -   i \Gamma \omega~.
\label{GeneralRipplonSpectrum2}
\end{equation}
For the ripplons at the interface between $^3$He-A and $^3$He-B the
friction parameter $\Gamma>0$ depends on temperature and is proportional 
to $T^3$ at low $T$.  The important property of the added dissipative
term is that it introduces the reference  frame of the horizontal wall.
The
$\omega$-dependence of the dissipative term in Eq.
(\ref{GeneralRipplonSpectrum2}), which has no Doppler shift,  implies 
that this spectrum  is written in the frame of the wall. We may expect
that under some conditions this description is applicable to the normal
viscous liquid, where the phenomenological parameter $\Gamma$ is
determined by Reynolds number of the flowing liquid and probably depends
on $\omega$ and $k$.

\section{Instability in the ergoregion}
\label{Instability}

If the non-zero $\Gamma$ is taken into account, from the spectrum 
$\omega(k)$ in Eq. (\ref{GeneralRipplonSpectrum2}) it follows that the
instability to the formation of the surface waves occurs when the
velocity $v$ of the flow with respect to the wall exceeds the critical
velocity $v_L$.  At $v=v_L$   the imaginary part ${\rm Im}~\omega(k_c)$
of the energy spectrum of the critical ripplon with momentum $k_c$
crosses zero and becomes positive, i.e. the attenuation of ripplons at
$v<v_L$ due  to dissipation  transforms to amplification at $v>v_L$
\cite{VolovikBook,Volovik}. The critical velocity $v_L$ and the momentum
of the critical ripplon $k_c$ do not depend on the friction parameter
$\Gamma$. They are different  in the ``relativistic'' and
``non-relativistic'' regimes:
\begin{eqnarray}
v_L=c~~,~~k_c=0~~,~~ ~~{\rm if}~~ hk_0 < \sqrt{3}~,
\label{CritVelRel}\\
v_L= c\sqrt{2/hk_0}~~,~~k_c=k_0 ~~,~~{\rm if}~~ hk_0 \gg 1~.
\label{CritVelNonRel}
\end{eqnarray}
In both regimes the frequency of the critical ripplon is $\omega(k_c)=0$, 
i.e. the critical ripplon must be stationary in the wall frame.

The fact that the threshold velocity $v_L$ does not depend on   $\Gamma$
demonstrates that the main role of the dissipative term is to provide the
reference frame of the wall with which the liquid interacts.  The flow of
a superfluid liquid with respect to this reference frame does not
experience any dissipation if its velocity is below $v_L$. The
dissipation  starts above the instability threshold when the surface of
the liquid is perturbed, i.e. ripplons are radiated due interaction of
the liquid with the wall. This indicates that the critical velocity of
the flow with respect to the wall coincides with the Landau criterion for
ripplon nucleation: 
\begin{equation}
v_L ={\rm min}_k \frac{E(k)}{k} ~~,~~E(k)=\sqrt{(\rho g +\sigma
k^2)/M(k)} ~.
\label{LandauVel}
\end{equation}
In the case of the interface between $^3$He-A and $^3$He-B, the
critical velocity of instability towards the growth of critical  ripplon
has been measured in the nonrelativistic  deep-water regime $hk_0\gg 1$
\cite{Kelvin-HelmholtzInstabilitySuperfluids}, and has been found in a
good agreement with  the theoretical estmate of the Landau velocity
(modified for the case of two liquids \cite{VolovikBook,Volovik}) wihout
any fitting parameter.

The region, where the flow velocity
$v$ exceeds  $v_L$, represents the ergoregion, since in the wall  frame
the energy of the critical ripplon is negative in this region, $E(k)+{\bf
k}\cdot {\bf v} <0$. For the relativistic ripplons, the ergoregion -- the
region where $v$ exceeds  $c$ -- is expressed in terms of the effective
metric in Eq.(\ref{EffectiveMetricCovariant}): in the ergoregion the
metric element $g_{00}$ changes sign and becomes positive. If the flow is
perpendicular to the  ergosurface (the boundary of the ergoregion), then
the ergosurface serves as the event horizon for ripplons. It is the black
hole horizon, if the liquid moves into the ergoregion, since ripplons
cannot excape from the ergoregion (if the non-relativistic $k^4$
corrections to the spectrum are ignored). Correspondingly,  if the liquid
moves from the ergoregion, the boundary of the ergoregion represents the
white hole horizon. 

The discussed instability of
the flow towards formation of ripplons in the supercritical region does
not depend on whether the horizon is of a black hole or of a white hole. 
This mechanism also does not resolve between the ergosurface and horizon.
The instability comes from the interaction with the fixed reference frame
and occurs in the region where the energy of the critical fluctuation is
negative in this frame. Such kind of instability is also called the Miles
instability
\cite{SchutzholdUnruh2}. In principle, Miles instability may take  place
behind the horizon of the atsronomical black holes  if there exists the
fundamental reference frame related for example with Planck physics
\cite{VolovikBook,SchutzholdUnruh2}. It  may lead to the decay of the
black hole much faster than the decay due to Hawking radiation.

\section{White-hole horizon in hydraulic jump}
\label{WH}

The situation with a white hole horizon is achieved in the so-called
hydraulic jump first discussed by Rayleigh in terms of the shock  wave
\cite{Rayleigh}. The circular hydraulic jump occurs when the vertical jet
of liquid falls on a flat horizontal surface. The flow of the liquid at
the surface exhibits a ring discontinuity at a certain distance $r=R$
from the jet  (observation of the non-circular hydraulic  jumps with
sharp corners has been reported in Ref. \cite{Ellegaard1}). At $r=R$
there is an abrupt increase in the depth $h$ of the liquid (typically by
order of magnitude) and correspondingly a decrease in the radial velocity
of the liquid. The velocity of the liquid in the interior region ($r<R$)
exceeds the speed of `light'  for ripplons
$v>c=\sqrt{hg}$, while outside the hydraulic jump ($r>R$) one has
$v<c=\sqrt{hg}$. Since the velocity flow is radial and outward, the
interior  region imitates the `white-hole' region.  The  interval of the
2+1 dimensional effective space-time in which the ``relativistic'' 
ripplons ``live''  is
\begin{equation}
ds^2=-c^2dt^2  + (dr-v(r)dt)^2 +r^2d\phi^2~.
\label{Interval} 
\end{equation}

The similar 3+1 dimensional space-time   in general relativity, the
so-called Painlev\'e-Gullstrand metric \cite{Painleve}
\begin{equation}
ds^2=-c^2dt^2  + (dr-v(r)dt)^2  +r^2\left( d\theta^2 + \sin^2\theta
\,d\phi^2 \right)~,~v^2(r)=\frac{2G M}{r},
\label{IntervalGR} 
\end{equation} 
is popular now in the black hole  physics (see Refs. \cite{RiverModel} 
and references therein).

In general relativity the metric  is continuous across the horizon. In
our case there is a real physical singularity at the white-hole horizon
-- the jump in the effective metric (\ref{EffectiveMetricCovariant}). 
However, the discussed mechanism of the Miles instability in the
ergoregion (or behind the horizon) does not depend on whether  the
horizon/ergosurface  is smooth or singular. 

A similar condensed matter analog of the black-hole horizon
with the physical singularity at the horizon has been
also discussed by Vachaspati \cite{Vachaspati}. At the boundary between
two superfluids the speed of sound (and thus the acoustic metric) has a
jump, $c_1\neq c_2$.   The acoustic black or white horizon occurs if the
superfluid velocity of the flow through the phase boundary is
subsonic in one of the superfluids but supersonic
in the other one, $c_1<v<c_2$. 

\section{Hydraulic jump in superfluids}
\label{HJ}

The analogy between the instability of the  surface inside the hydraulic
jump and the instability of the vacuum behind the horizon can be useful
only if the liquid simulates the quantum vacuum. For that, the liquid
must be quantum, and its flow should not exhibit any friction in the
absence of a horizon. That is why the full analogy could occur if one
uses either the flow of quantum liquid with high Reynolds number, or the
superfluid liquid which has no viscosity. Quantum liquids such as
superfluid or normal $^3$He and $^4$He are good candidates.

The first observation of the circular hydraulic  jump in superfluid
liquid (superfluid $^4$He) was reported in Ref. \cite{Pettersen}. The
surface waves generated in the ergoregion (in the region inside the jump)
were observed. The critical ripplon appeared to be stationary in the wall
frame in agreement with the Miles instability towards ripplon radiation
inside the ergoregion discussed in Sec. \ref{Instability}.  This is the
first experiment, where the analog of the instability of the vacuum
inside the horizon has been simulated. The growths of the critical
ripplon is saturated due to the non-linear effects, and then the whole
pattern remains stationary (though not static). This is different from
the case of the instability observed at the interface between $^3$He-A
and $^3$He-B, where the instability is not saturated and leads to the
crucial rearrangement of the vacuum state: Quantized vortices penetrate
into the  $^3$He-B side from $^3$He-A, they partially screen the
$^3$He-B  flow and reduce its velocity back below the threshold  for the
ripplon formation
\cite{Kelvin-HelmholtzInstabilitySuperfluids}.

Under the conditions of experiment   \cite{Pettersen} the hydraulic jump
in superfluid $^4$He is very similar to that in the normal liquid $^4$He.
The position $R$ of the hydraulic jump as a function of temperature  does
not experience discontinuity at the superfluid transition. This suggests 
that quantized vortices are formed, which provide the mutual friction
between the superfluid and normal components. As a result even below the
$\lambda$-point,  the liquid moves as a whole though with lower viscosity
because of the reduced fraction of the normal component.

To avoid the effect of the normal component  it would be desirable to
reduce the temperature or to conduct similar experiments in a shallow
superfluid $^3$He. 

The advantage of superfluid $^3$He is that,  as distinct from the
superfluid $^4$He, vortices are not easily formed there:  the energy
barrier for vortex nucleation in $^3$He-B is about
$10^6$ times bigger than temperature \cite{Parts}.  In addition, in
superfluid $^3$He  the normal component of the liquid is very viscous
compared to that in superfluid $^4$He. In the normal state the kinematic
viscosity is $\nu\sim 10^{-4}$ cm$^2$/s in liquid $^4$He, and $\nu\sim 1$
cm$^2$/s in liquid $^3$He.    That is why in  many practical arrangements
the normal component in superfluid $^3$He remains at rest with respect to
the reference frame of the wall and thus does not produce any
dissipation  if the flow of the superfluid component is sub-critical. One
can also exploit thin films of a superfluid liquid, where the normal
component is fixed. The ripplons there represent the so-called third
sound (recent discussion on the third sound propagating in superfluid
$^3$He  films can be found in Ref. \cite{Sauls}).  In 1999 Seamus Davis
suggested to use the third sound in superfluid $^3$He for simulation of
the horizons \cite{Davis}.

In normal liquids it is the viscosity which determines  the position $R$
of the hydraulic jump (see \cite{Bohr,Ray}). The open question is what is
the dissipation mechanism which determines the position $R$ of the
white-hole horizon in a superfluid flow with stationary or absent normal
component when its viscosity is effectively switched off.  Since there is
no dissipation of the superfluid flow if its velocity is below $v_L$, one
may expect that  the same mechanism, which is responsible for dissipation
in the presence of the horizon, also determines the position $R$ of the
horizon. If so, the measurement of $R$ as function of parameters of the
system will give the information on various mechanisms of decay of white
hole. If the Miles vacuum instability towards ripplon radiation inside
the horizon  is saturated as in experiment  \cite{Pettersen}, the other
mechanisms will intervene such as the black-hole laser \cite {BHlaser},
and even the quantum mechanical Hawking radiation of ripplons. The latter
should be enhanced at the sharp discontinuous horizon of the hydraulic
jump and maybe near the sharp corners of the non-circular (polygonal) 
hydraulic jump observed in Ref.  \cite{Ellegaard1}. 

  It is also unclear whether  it is possible  to approach the limit of a 
smooth horizon,  without the shock wave of the hydraulic jump; and
whether it is possible to construct the inward flow of the liquid which
would serve as analog of the black hole horizon. 

I  thank Etienne Rolley and Michael Pettersen  who sent me their
experimental results prior to publication and Marc Rabaud and Tomas Bohr
for discussions. This work is supported in part by the Russian Ministry
of Education and Science, through the Leading Scientific School grant
$\#$2338.2003.2, and by the European Science Foundation  COSLAB Program.

\end{document}